%% file: JT2015c.tex
\documentclass[dvipsnames]{aa}
\pdfoutput=1

\ifx\pdfoutput\undefined
\usepackage{graphicx}
\else
\usepackage[pdfauthor={J.~Threlfall},citecolor=blue,linkcolor=blue, linktocpage=true,colorlinks=true]{hyperref}
\usepackage{epstopdf}
\fi 

\usepackage{natbib}
\bibpunct{(}{)}{;}{a}{}{,} 
\usepackage{txfonts}
\usepackage{subfig}	
\usepackage{epsfig}	
\usepackage{cancel}
\usepackage{multirow}
\usepackage{fixltx2e}	
\usepackage{amssymb}	
\graphicspath{{figs/}}

\newsavebox{\bigleftbox}
\newsavebox{\bigrightbox}
\usepackage{color}
\definecolor{darkgreen}{rgb}{0,0.45,0}

\DeclareRobustCommand*{\unit}[1]{\def~{\,}\ensuremath{\mathrm{\,#1}}}

\input{math_def.tex}	
\date{}			

\begin{document}
 \title{Particle acceleration at reconnecting separator current layers}
 \titlerunning{Particle acceleration at reconnecting separator current layers}
 \author{J.~Threlfall \and  J.~E.~H.~Stevenson \and C.~E.~Parnell \and T.~Neukirch }
 \institute{School of Mathematics and Statistics, University of St Andrews, St Andrews, Fife, KY16 9SS, U.K. \email{\{jwt9;jm686;cep;tn3\}@st-andrews.ac.uk}}
 \abstract
 {}
 {The aim of this work is to investigate and characterise particle behaviour in a 3D magnetohydrodynamic (MHD) model of a reconnecting magnetic separator.}
 {We use a relativistic guiding-centre test-particle code to investigate electron and proton acceleration in snapshots from 3D MHD separator reconnection experiments, and compare the results with findings from an analytical separator reconnection model studied in a previous investigation.}  
 {The behaviour and acceleration of large distributions of particles are examined in detail for both analytical and numerical separator reconnection models. Accelerated particle orbit trajectories are shown to follow the separator before leaving the system along the separatrix surface of one of the nulls (determined by particle species) in the system of both models. A sufficiently localised electric field about the separator causes the orbits to appear to follow the spine bounding the separatrix surface field lines instead. We analyse and discuss the locations and spread of accelerated particle orbit final positions, which are seen to change over time in the numerical separator reconnection model. We deduce a simple relationship between the final energy range of particle orbits and the model dimensions, and discuss its implications for observed magnetic separators in the solar corona.}
 {}
 \maketitle

\section{Introduction}\label{sec:Intro}
Particle acceleration is one of the most commonly observed and yet possibly least understood phenomena recorded during a solar flare.
The mechanism through which electrons/protons may be rapidly accelerated to tens of MeV/several GeV is one of the biggest unsolved problems in solar physics to date \citep[see e.g.][and references therein]{review:Fletcheretal2011,review:Cargilletal2012}.
It is generally accepted that magnetic reconnection plays a role on some level in the acceleration process.

Magnetic reconnection is a fundamental mechanism in plasma physics, both in space and in the laboratory. It lies at the heart of many dynamic solar, magnetospheric, and astrophysical phenomena. It is a process by which local and global magnetic fields restructure into more energetically favourable configurations converting considerable amounts of magnetic energy into bulk plasma motions, thermal energy, as well as accelerating particles \citep[e.g.][]{book:PriestForbes, book:BirnPriest}. As shown, for example, by \cite{paper:Schindleretal1988,paper:HesseSchindler1988,paper:Schindleretal1991} it is generically associated with parallel electric fields; these parallel electric fields are likely candidates to accelerate particles.

Macroscopic (e.g. magnetohydrodynamic or MHD) models of reconnection events are insufficient to model plasma behaviour at very small (microscopic) scales. Plasma behaviour at these scales can be uncovered using, for example, a test particle approach. While early test particle studies focussed on acceleration in 2D reconnection models with and without guide fields for simplicity, more recent studies are now able to consider particle behaviour in fully three-dimensional (3D) reconnection experiments \citep[a more comprehensive discussion of reconnection models of particle behaviour in various dimensions can be found in e.g.][]{paper:Threlfalletal2015}. Of these more recent 3D experiments, some investigate particle acceleration associated with single isolated topological features, whilst others study acceleration in more complex scenarios where the reconnection takes place at multiple sites within the modelled large-scale structures. One of these topological features, a magnetic separator, has (until recently) been overlooked in regard to their capacity as a potential particle accelerator, despite evidence of their importance in various magnetic interaction events, such as flux emergence \citep[e.g.][]{paper:Parnelletal2010b}, which is often associated with solar flares.

Magnetic separators are single magnetic field lines which link pairs of magnetic null points; these field lines lie at the intersection of four topologically distinct flux domains. In so doing, they may be loosely regarded as the 3D analogue of a 2D X-point or X-point plus guide-field configuration. Separators are likely sites of current sheet formation \citep{paper:LauFinn1990, paper:Haynesetal2007,paper:Parnelletal2010a,paper:Stevensonetal2015}, and provide unique conditions for the accumulation of high current density on a large scale. Unlike 3D null points, magnetic separators have the potential to traverse vast distances within the solar corona \citep{paper:Closeetal2004,paper:Parnelletal2010b,paper:Plattenetal2014}; at these locations magnetic reconnection would no longer be constrained in highly localised regions (as it is at 3D null points), but may take place across extended regions. 

Separator reconnection has been observed in the solar corona \citep{paper:Longcopeetal2005} and its presence has been confirmed within Earth's magnetosphere through in situ observations \citep[e.g.][]{paper:Xiaoetal2007,paper:Dengetal2009,paper:Guoetal2013} and also numerical models \citep[e.g.][]{paper:DorelliBhattacharjee2008,paper:Komaretal2013}. Observational support for particle acceleration models based on separator reconnection events can also be found in the literature \citep[e.g.][]{paper:Metcalfetal2003}. Despite this, only a single attempt has been made thus far to model particle behaviour in the vicinity of a reconnecting magnetic separator. In this work, \citet{paper:Threlfalletal2015} were able to show that the parallel electric field generated along a magnetic separator is sufficient to rapidly accelerate test particles to non-thermal energies. For this first investigation of particle behaviour in a separator reconnection event, \citet{paper:Threlfalletal2015} used a simple analytical kinematic model. Our goal is to extend this initial investigation, by studying particle behaviour in MHD experiments of current sheet formation \citep{paper:Stevensonetal2015} and reconnection \citep{inpress:StevensonParnell2015b} at magnetic separators.

\begin{figure}[t]
\includegraphics[width=0.49\textwidth]{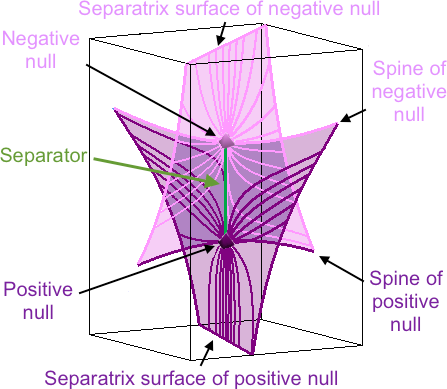}	
\caption{Illustration of the topological features in the initial magnetic field of the separator reconnection experiments. Both experiments considered here (MHS1 and MHS2) have two separatrix surfaces which intersect at a single magnetic field line, called a separator, linking oppositely-signed magnetic null points.}
\label{fig:sepcartoon}
\end{figure}
The primary objective of the present work is to determine how particles behave in the vicinity of a typical separator reconnection event. To do this, we compare orbits based on a simple analytical model of separator reconnection \citep[detailed in][]{paper:Threlfalletal2015} with orbits obtained from a recent detailed study of a numerical separator reconnection event, where a separator (initially in magnetohydrostatic, or MHS, equilibrium) is allowed to reconnect \citep{inpress:StevensonParnell2015b}.
The paper is organised as follows; In Sect.~\ref{sec:model} we discuss our approach to the problem, introducing both the global environment of the MHD experiments (described in Sect.~\ref{subsec:MHDmodel}) and the equations which govern test particle motion (in Sect.~\ref{subsec:GCmodel}). 
General comparisons between the analytical and numerical separator reconnection model snapshots and their impact on particle behaviour are discussed in Sect.~\ref{sec:an} before an examination of changes in orbit behaviour over the course of the reconnection event is presented in Sect.~\ref{sec:time}. A discussion of our findings is presented in Sect.~\ref{sec:disc} followed by our conclusions and future areas of study which are outlined in Sect.~\ref{sec:conc}.

\section{Model setup}\label{sec:model}
Our model can be broadly split into two parts: a snapshot taken from a time-dependent MHD experiment (modelling the dissipation of current held in MHS equilibrium and accumulated along a separator) into which we insert particles, and the test particle motion itself. A brief overview of both parts are described in the following sections:

\subsection{MHD separator reconnection model}\label{subsec:MHDmodel}
In this paper, we consider MHD separator reconnection experiments starting from two MHS equilibrium fields, which were formed in an identical manner to that detailed in \citet{paper:Stevensonetal2015}, i.e. through the non-resistive relaxation of an analytical magnetic field using the MHD code Lare3d \citep{paper:LareXd2001}. This analytical field consists of two oppositely-signed 3D null points whose separatrix surfaces intersect to form a separator, which lies along the $z$-axis linking the nulls and is similar to the configuration shown in Fig.~\ref{fig:sepcartoon}. 

Before the non-resistive relaxation begins, the plasma pressure and current are uniform with the current directed parallel to the separator along the $z$-axis. However, since the analytical field is not in force-balance, as soon as the non-resistive relaxation experiment begins the Lorentz force causes the separatrix surfaces of the nulls to collapse towards each other. This leads to a strong twisted current layer forming along the separator and regions of enhanced/diminished plasma pressure forming within/outwith cusps about the separator. As the relaxation continues a plasma pressure force acts to counter-balance this collapse until the Lorentz and plasma pressure forces balance. A MHS equilibrium is achieved everywhere in a finite time, except very locally about the separator (where the current layer exists) and very close to the separatrix surfaces (where the current is enhanced); in these regions due to an infinite time singularity very small residual forces remain \citep[see][for details]{paper:Stevensonetal2015}. We note that in this work we use the phrase "separatrix surface" and "fan plane" interchangeably.

In the MHD reconnection experiments studied here, reconnection is triggered in the current layers of the MHS equilibrium by the addition of an anomalous diffusivity, $\eta$($=1/\mu_0\sigma$, for an electrical resistivity $1/\sigma$ and permeability of free space $\mu_0$), in the governing MHD equations. The resistive reconnection experiments were carried out using Lare3d \citep[and are studied in detail in][]{inpress:StevensonParnell2015b}. The value of the diffusivity is zero unless the current is above a given threshold, $j_{\rm{crit}}$. The value of $j_{\rm{crit}}$ is chosen such that only the strong current within the separator current layer (and not the enhanced current on the separatrix surfaces of the nulls) is involved in the reconnection. 

Since the electric field is defined, using Ohm's Law, as 
\[
{\bf{E}}=\frac{1}{\sigma}{\bf{j}}-{\bf{v}}\times{\bf{B}},
\]
parallel electric fields (i.e. $E_{||}={\bf{E}}\cdot{\bf{B}}/|{B}| = {\bf j}\cdot{\bf B}/\sigma|{B}|$) can only exist in regions of non-zero finite resistivity $1/\sigma$. Thus, in our models, $E_{||}$ is only generated by regions of current density at or above the critical current threshold $j_{\rm{crit}}$. As reconnection takes place, the amount of current density at or above this threshold decreases, until all the remaining current density in the system is below $j_{\rm{crit}}$, at which point reconnection effectively ceases (see discussion in Sect.~\ref{sec:time}). Changing the value of $j_{\rm{crit}}$ will not change the peak electric field generated by resistive effects (unless $j_{\rm{crit}}$ exceeds the peak current, whereupon $E_{||}=0$). Instead, changing $j_{\rm{crit}}$ will alter the minimum value of resistively generated electric field, which ultimately determines the size of the reconnection region volume, the total flux reconnected and also the duration of reconnection \citep[for further details on the nature of the reconnection and how various parameters affect it, see][]{inpress:StevensonParnell2015b}.

The two MHS equilibrium fields from which we start our MHD reconnection experiments (MHS1 and MHS2) have a similar topological make up to that discussed in \citet{paper:Stevensonetal2015} and are formed in the same way. Each contains a twisted current layer lying along the separator which has equilibrium length $l_{\rm{sep}}$. These two fields are very similar; the only differences are noted in Table~\ref{table:sepdata}. Specifically, MHS1 has a low plasma beta away from the nulls and at the separator, which runs along the $z$-axis linking the two nulls that lie a distance of almost 3 apart. In MHS2, the plasma beta is high, all along the single shorter separator.
\footnote{In this context, a "low" plasma beta value means $\beta<1$, while a "high" plasma beta value means $\beta>1$; the value of $\beta=1$ represents a transition from a plasma dominated by thermal effects to one dominated by magnetic effects. While the value of $\beta$ used in MHS1 ($\beta=0.77$) does not seem particularly low in relation to modelling solar coronal plasma, in reality it is difficult to form MHS equilibrium fields containing separator current layers when $\beta<<1$ \citep[as discussed in e.g.][]{thesis:Stevenson}.}
 
Both MHS fields have dimensionless domain size $-1 \le x,y \le 1$ with a grid resolution of $(512,512,768)$. Particle acceleration in a single snapshot of the reconnection model starting with MHS1 is discussed in Sect.~\ref{subsec:compy}. The nature and consequences of the reconnection which occurs in the model starting from MHS2 is detailed in \citep{inpress:StevensonParnell2015b} and the particle acceleration determined in various snapshots throughout this reconnection experiment are discussed in Sect.~\ref{sec:time}.  
\begin{table}[ht]
\centering
\caption{Non-dimensional values for two initial MHS equilibria (MHS1 and MHS2) with separator lengths $l_{\rm{sep}}$ along the $z$-axis, the mean value of the plasma beta, $\bar{\beta}$, the plasma beta half-way along the separator, $\beta_{l_{\rm{sep}}/2}$, the peak value of the current along the main separator and the value of $j_{\rm{crit}}$ used in the reconnection experiment.}
\label{table:sepdata}
\begin{tabular}{c|c|c}
Parameter & MHS1  & MHS2 \\ \hline
$l_{\rm{sep}}$ & 2.87  & 1.18 \\ 
$z$ dimensions &  $-1 \le z \le 4$ & $-1.75 \le z \le 2.75$ \\
$\bar{\beta}$ & 0.77 & 4.83 \\
$\beta_{l_{\rm{sep}}/2}$ & 0.72 & 115.50 \\
Peak $|{\bf{j}}|$ along sep. & 26.19 & 22.40 \\
$j_{\rm{crit}}$ & 18 & 10                    
\end{tabular}
\end{table}

\subsection{Relativistic particle dynamics}\label{subsec:GCmodel}
Having established the separator reconnection environment which we will study, all that remains is to outline the equations which will govern particle behaviour. The approach we use is identical to that of \citet{inpress:Threlfalletal2015b}, which was applied to the study of particle behaviour in MHD simulations of a non-flaring active region and also \citet{paper:Threlfalletal2015}, which investigated acceleration in an analytical separator reconnection model. Similar test-particle studies using MHD experiments with different reconnecting magnetic configurations have also been performed \citep[by][for example]{paper:GordovskyyBrowning2011,paper:Gordovskyyetal2014}. 

Very briefly, a particle with rest-mass $m_0$, charge $q$ and relativistic magnetic moment, $\mu_r$, will move around its guiding centre, located at ${\bf{R}}$, subject to a magnetic field ${\bf{B}}$ (with magnitude $B(=|{\bf{B}}|)$ and unit vector ${\bf{b}}(={\bf{B}}/B)$) and an electric field ${\bf{E}}$ according to the following equations \citep{book:Northrop1963}:
\begin{subequations}
 \begin{align}
  \frac{d{u_\parallel}}{dt}&=\frac{d}{dt}\left(\gamma\vpar\right)=\gamma\ue\cdot{\frac{d{\bf{b}}}{dt}}+\omscl\tscl E_\parallel-\frac{\mu_r}{\gamma}\frac{\partial{B^\star}}{\partial s}, \label{eq:Rnorm1} \\
  {\bf\dot{R}_\perp}&=\ue+\frac{\bf{b}}{B^{\star\star}}\times\left\lbrace \frac{{1}}{\omscl\tscl}\left[ \frac{\mu_r}{\gamma}\left( \grad{B^\star}+ \frac{\vsclsq}{{c^2}}\ue\frac{\partial B^\star}{\partial t}\right)\right.\right. \nonumber \\ 
   &\qquad\quad\qquad\qquad\left.\left. +u_\parallel\frac{d{\bf{b}}}{dt}+\gamma\frac{d\ue}{dt}\right]+\frac{\vsclsq}{{c^2}}\frac{u_\parallel}{\gamma}{E_\parallel}\ue \right\rbrace, \label{eq:Rnorm2} \\ 
  \frac{d\gamma}{dt}&=\frac{\vsclsq}{{c^2}}\left[\omscl\tscl\left({\bf\dot{R}_\perp}+\frac{u_\parallel}{\gamma}{\bf{b}}\right)\cdot{\bf{E}}+\frac{\mu_r}{\gamma}\frac{\partial B^\star}{\partial t}\right],   \label{eq:Rnorm3} \\
  \mu_r&=\frac{\gamma^2{\vperp^2}}{B}, \label{eq:Rnorm4}  
 \end{align}
 \label{eq:rel_norm} 
\end{subequations}
where, for a given magnetic field strength $B$, $B^{\star}$ and $B^{\star\star}$ are defined as
\[
 B^\star=B\left( 1-\frac{1}{c^2}\frac{{\Eperp}^2}{B^2}\right)^{\frac{1}{2}} , \qquad B^{\star\star}=B\left(1-\frac{1}{c^2}\frac{{\Eperp}^2}{B^2}\right),
\]
and also noting that $B^\star$ and $B^{\star\star}$ retain the dimensions of $B$ (since the multiplying quantities are dimensionless).

The particle orbit behaviour is particularly affected by guiding centre drifts. The majority of behaviour changes are often caused by the ${E}\times{B}$ drift, which causes the orbit to drift at a velocity $\ue(={\bf{E}}\times{\bf{b}}/B)$. The components of velocity and electric field that are aligned with the magnetic field (from now on known as the parallel velocity and parallel electric field) are $\vpar(={\bf{b}}\cdot{\dot{\bf{R}}})$ and $\Epar(={\bf{b}}\cdot{\bf{E}})$, respectively. $\vperp$ is the gyro-velocity, $\dot{\bf{R}}_\perp(=\dot{\bf{R}}-\vpar{\bf{b}})$ is the perpendicular component of guiding centre velocity and $s$ is an arc-length along the magnetic field-lines. Finally, for simplicity of notation a relativistic parallel velocity $\upar(=\gamma\vpar)$, where $\gamma \left(= c/\left(c^2-v^2\right)^{1/2}\right)$ is the Lorentz factor.

In this work we consider either electrons or protons thus, for electrons, the rest mass and charge are $m_0=m_e=9.1\times10^{-31}$\unit{kg} and $q=e=-1.6022\times10^{-19}$\unit{C}, respectively. For protons the rest mass and charge are $m_0=m_p=1.67\times10^{-27}$\unit{kg} and $q=|e|=1.6022\times10^{-19}$\unit{C}. 

Eqs.~(\ref{eq:rel_norm}) have been non-dimensionalised. To recover dimensional values, appropriate dimensional parameters are applied. Since this work is motivated by the behaviour of particles in a solar coronal environment, for all the MHD experiments the normalising parameters for the magnetic field, length and time are $\bscl=10$\unit{G}, $\lscl=1$\unit{Mm}  and $\tscl=20$\unit{s}, respectively. Thus, velocities in the model are scaled by $\vscl(={\lscl}{\tscl}^{-1}) = 5\times10^4$\unit{ms}$^{-1}$, electric fields by $\escl(={\bscl\,\lscl}{\tscl}^{-1}=\bscl\,\vscl)=50$\unit{V m}$^{-1}$ and energies by $\KEscl(=m_0 c^2 \times {\vscl}^2/2c^2)=511\times 1.388\times10^{-8}\unit{keV}$ for electrons and $938\times 1.388\times10^{-8}\unit{MeV}$ for protons. An additional normalising constant is, $\omscl(={q\,\bscl}{m_0}^{-1})$, where the factor $\omscl\tscl$ controls the significance of particular terms at particular scales. For comparison with earlier investigations \citep{paper:Threlfalletal2015}, the normalising parameters for the analytical separator model used in Sect.~\ref{subsec:scaling} differ from above. 

In this investigation, we will (where possible) use a scale-free form in any discussion of results. A description of how this scale-free form is derived can be found in Sect.~\ref{sec:an}.

In order to determine the particle behaviour, Eqs.~(\ref{eq:rel_norm}) are evolved in time using a 4th order Runge-Kutta scheme with a variable timestep, according to the electric and magnetic fields found in a single snapshot of the particular numerical MHD experiment. In Sect.~\ref{subsec:scaling} the MHD experiment considered is that which starts from the initial equilibrium MHS1, whilst in Sect.~\ref{subsec:compy} the MHD snapshots come from the experiment starting with MHS2. Details of the MHD behaviour of these experiments are outlined in \cite{inpress:StevensonParnell2015b}.

Using the test particle approach in this way requires that the gyro-radius and gyro-period of each orbit must remain well below the normalising length-scales and time-scales of the macroscopic MHD environment. The validity of this assumption is tested and discussed in Sec.~\ref{sec:an}. 

\section{Comparison with analytical separator model}\label{sec:an}
In order to build up a general picture of particle orbit behaviour in the vicinity of a reconnecting magnetic separator, we first compare orbits found in the numerical MHD experiments of separator reconnection with orbits recovered from an analytical model of separator reconnection.
\begin{figure*}
 \centering
 \sbox{\bigleftbox}{%
 \begin{minipage}[b]{.49\textwidth}
  \centering
  \vspace*{\fill}
  \subfloat[electron initial positions]
  {\label{subfig:JT2015einipos}\resizebox{\textwidth}{!}{\includegraphics{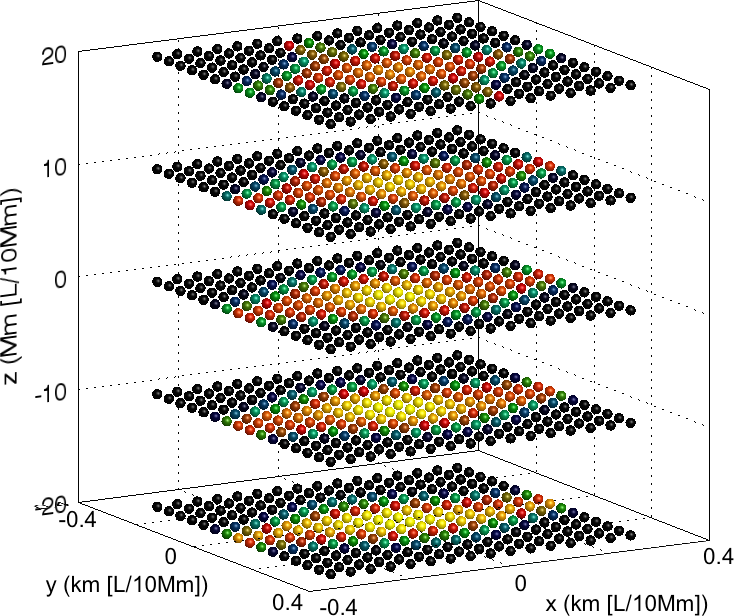}}}\\
  \vspace*{1em}
  \renewcommand{\thesubfigure}{c}
  {\resizebox{\textwidth}{!}{\includegraphics{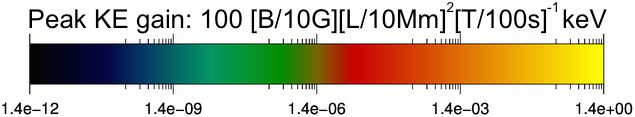}}}\\
  \subfloat[proton initial positions]{\label{subfig:JT2015pinipos}\resizebox{\textwidth}{!}{\includegraphics{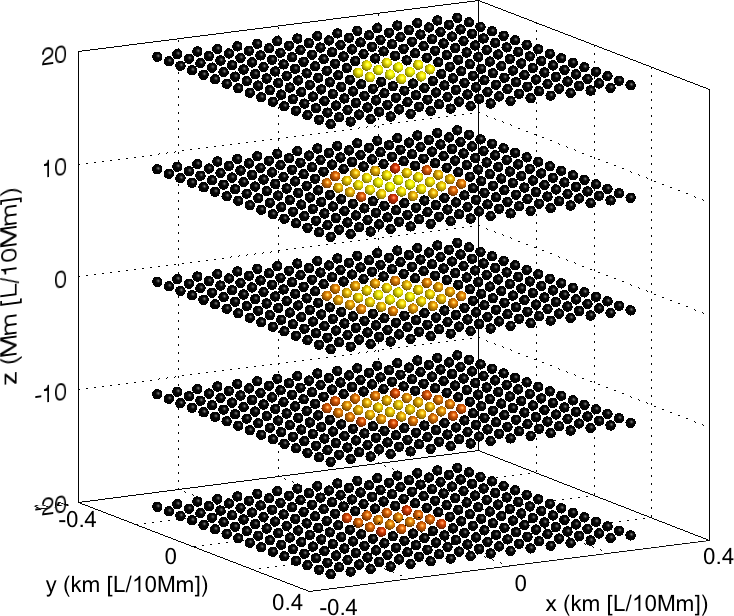}}}
 \end{minipage}%
 }\usebox{\bigleftbox}%
 \begin{minipage}[b][\ht\bigleftbox][s]{.445\textwidth}
  \centering
    \renewcommand{\thesubfigure}{b}
   \subfloat[electron final positions] {\label{subfig:JT2015eorbs}\resizebox{\textwidth}{!}{\includegraphics{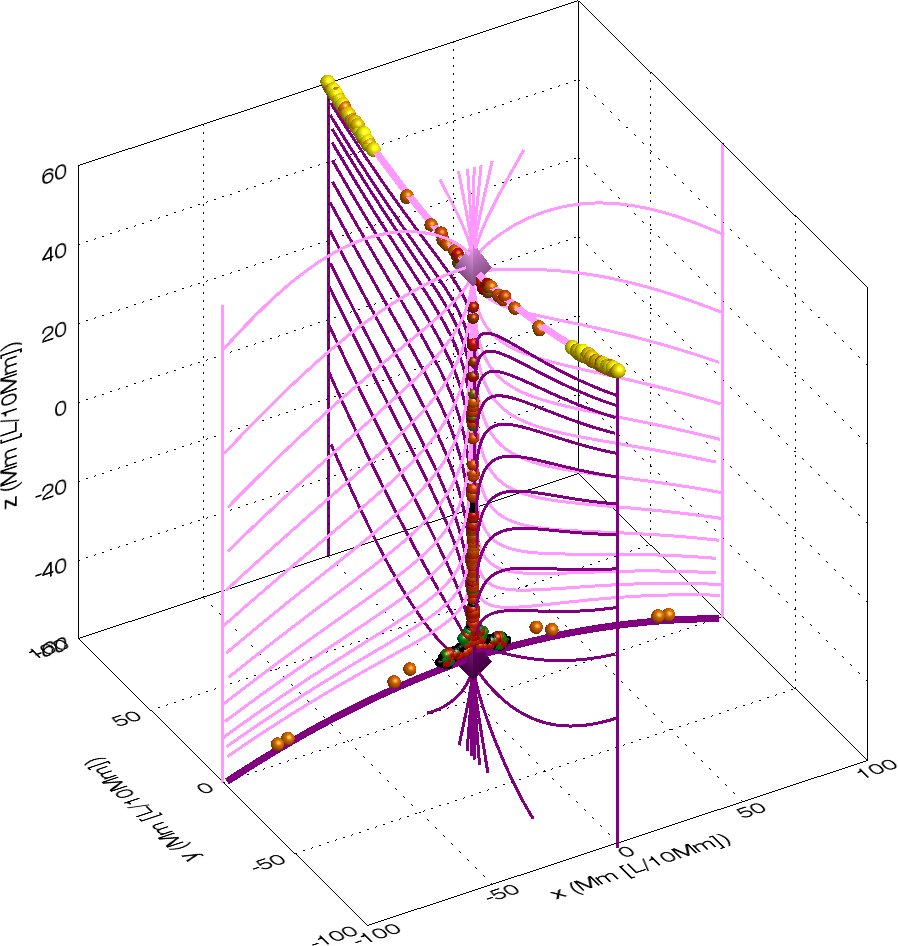}}}\\
     \renewcommand{\thesubfigure}{d}
   \subfloat[proton final positions] {\label{subfig:JT2015porbs}\resizebox{\textwidth}{!}{\includegraphics{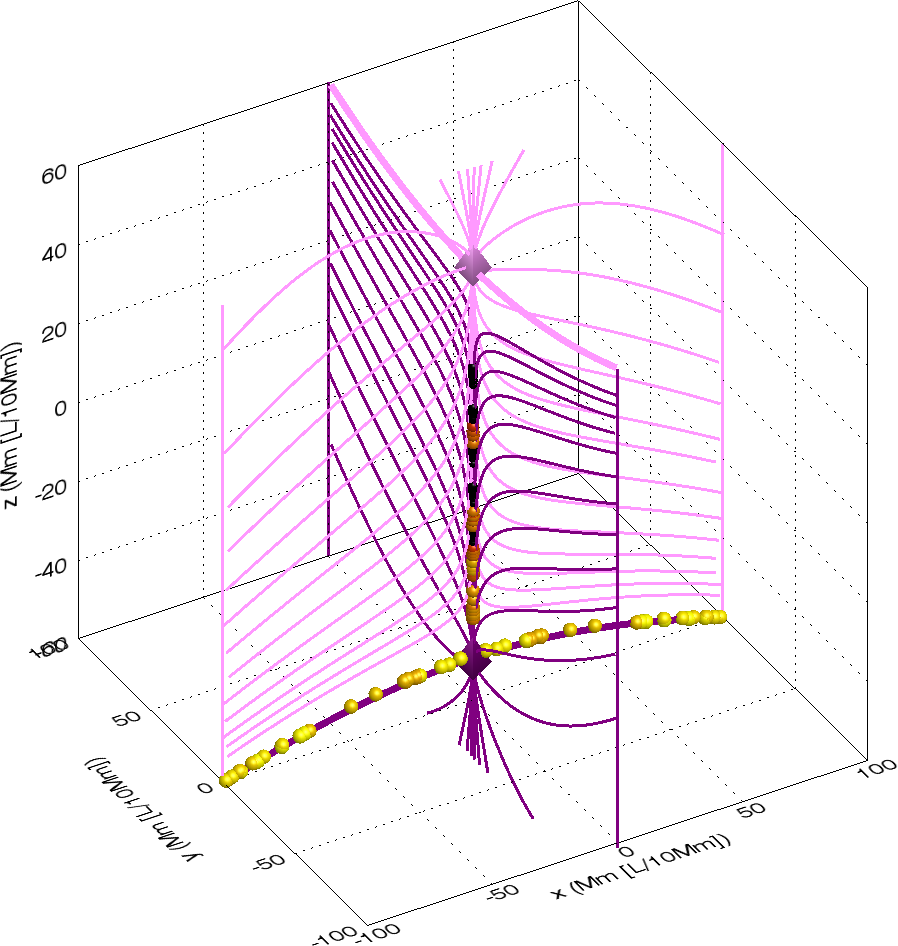}}}
 \end{minipage}
\caption{Analytical separator model \citep[][]{paper:Threlfalletal2015}; \protect\subref{subfig:JT2015einipos} initial positions (closeup around separator) and \protect\subref{subfig:JT2015eorbs} final positions 
for electrons with initial pitch angle $45\degr$ and initial kinetic energy of $20$\unit{eV}. \protect\subref{subfig:JT2015pinipos} and \protect\subref{subfig:JT2015porbs} are the same, but for protons. The colour of each particle identifies the (scaled) peak kinetic energy gained during each orbit (see colour bar). For \protect\subref{subfig:JT2015eorbs} and \protect\subref{subfig:JT2015porbs} thin pink/purple lines are field lines in the fan planes associated with the upper/lower nulls (indicated by trapezoids) and thick pink/purple lines indicate the corresponding null's spine. Also in these graphs, to aid visualisation, the orb size for each particle has been enhanced in proportion to its final energy. For reference, $\bscl=10$\unit{G}, $\lscl=10$\unit{Mm}, and $\tscl=100$\unit{s} for these orbit simulations.}
 \label{fig:JT2015}
\end{figure*} 

\subsection[]{Scale-free results of \citet{paper:Threlfalletal2015}}\label{subsec:scaling}
The work highlighted in \citet{paper:Threlfalletal2015} showed that it was possible even for a simple analytical model of separator reconnection to generate particle acceleration of electrons and protons to non-thermal energies. The analytical model used in this work \citep[originally devised by][]{paper:Wilmot-SmithHornig2011}, described a uniform magnetic separator field, surrounded by an electric field region which is  generated by a time-dependent ring of magnetic flux. In order to study particle acceleration, \citeauthor{paper:Threlfalletal2015} scaled the non-dimensional model of \citeauthor{paper:Wilmot-SmithHornig2011} using approximately solar coronal values and restricted the model parameters to obtain a reconnection region/current sheet width of appropriate size (approximately $10$ ion inertial lengths, $10c/\omega_{pi}$, where $\omega_{pi}$ is the local plasma frequency). However, as discussed in \citet{paper:Threlfalletal2015}, these results are entirely scalable; applying a different field strength, length-scale or time-scale to the model of \citet{paper:Wilmot-SmithHornig2011} would yield identical orbits, energy gains, etc, multiplied by some factor. In order to directly compare and contrast the analytical and numerical models independently of any applied scales, we first recast some of the results of \citet{paper:Threlfalletal2015}, in Fig.~\ref{fig:JT2015}, to highlight exactly how different scales might affect the resulting energy gains.

Figure~\ref{fig:JT2015} illustrates the scale independent results of the analytical model; for the chosen reconnection region width of 72\unit{m} and height ($20$\unit{Mm}) along a separator of length $100$\unit{Mm} with a field strength of $10$\unit{G} studied for $100$\unit{s}, we may recover peak particle energies of up to $140$\unit{keV}. 
Different applied scales using the same configuration should yield different energy ranges. These energy ranges can be recovered by scaling the results in Fig.~\ref{fig:JT2015}. 

The largest contributions to the energy gained in each orbit results from the work done by the parallel electric field. Despite the electric field resulting from an inductive magnetic field (rather than the gradient of an applied potential), this can still be interpreted as a form of potential difference \citep[see e.g.][]{paper:Schindleretal1991}. Further references to a "potential difference" refer to such a (field-aligned) potential, which results in an energy change of the form:
\[
\Delta=\int E_{||} ds=\frac{\lscl^2\bscl}{\tscl}\int{\tilde{E_{||}}d\tilde{s}}=\frac{\lscl^2\bscl}{\tscl}\tilde{\Delta},
\]
where the barred quantities represent the normalised equivalent of each variable. Changing the scale of the global environment in the experiment will rescale energy variations in this manner without requiring any recalculation of particle orbits. Thus, increasing the field strength by a factor of ten increases the peak energy range by that same factor and hence would recover the range found in \citet{paper:Threlfalletal2015} (they used a field strength of $100$\unit{G} instead of $10$\unit{G}, but all other parameters were identical to those used here and they recovered gains of up to $1.4$\unit{MeV}$=10\times140$\unit{keV}).

It is also clear that the choice of length-scale is critical in these experiments; the $\lscl^2$ dependence results from a combination of the definition of the electric field and the length over which this field extends. As another example, a $10$\unit{G} separator of length $10$\unit{Mm} (reconnection region length $2$\unit{Mm}) would yield a hundredfold decrease in peak particle energy seen in Fig.~\ref{fig:JT2015} (reducing the peak to $1.4$\unit{keV}). Thus, we can use the findings presented in Fig.~\ref{fig:JT2015} to broadly estimate particle orbit energy gains in a number of different separator reconnection cases, based on specific observational examples.

Briefly reminding ourselves of the primary findings of \citet{paper:Threlfalletal2015}, Fig.~\ref{fig:JT2015} demonstrates the locations and energies which can be achieved by acceleration caused by the parallel electric field generated in a simple kinematic separator reconnection model. Accelerated particle orbits are concentrated very close to the separator and are driven out along fan planes very close to the spine of one of the nulls. Both protons and electrons typically achieve large energy gains, particularly those which begin at opposite ends of the separator (shown by the yellow orbs in Figs.~\ref{subfig:JT2015einipos} \&~\ref{subfig:JT2015pinipos}) and leave the model close to opposing null spines (shown by the yellow orbs in Figs.~\ref{subfig:JT2015eorbs} \&~\ref{subfig:JT2015porbs}). Particle distributions strongly reflect the initial choice of electric field configuration.  However, in Fig.~\ref{subfig:JT2015einipos}, electrons originally outside regions of strong $E_{||}$ may also be moderately accelerated, if they orbit close to a fan plane surface which allows them to enter the reconnection region at later times.

\subsection{Scale-free numerical reconnection model results}\label{subsec:compy}
For comparison with the (now scale-free) analytical kinematic separator reconnection model, we first study a single snapshot from the numerical separator reconnection event starting from MHS1, at the time when the rate of reconnection is close to its peak value. We will compare the orbits described in Sec.~\ref{subsec:scaling} with orbits determined from this MHD snapshot.

The applied scalings for the MHD experiment are $\lscl=1$\unit{Mm}, $\bscl=10$\unit{G} and $\tscl=20$\unit{s}; the separator is
$2.87$\unit{Mm} long with nulls at $(0,0,0.07)$\unit{Mm} and $(0,0,2.94)$\unit{Mm}, in a simulation domain which extends from $-1\rightarrow4$\unit{Mm} in $z$ and $-1\rightarrow1$\unit{Mm} in $x$ \& $y$. The reconnection region is determined by a current threshold above which a diffusivity of $\eta=0.001$ is applied. The critical current value, $j_{\rm{crit}}$, was chosen to be large enough that current structures would be accurately resolved by the code and would only allow the diffusivity to act upon a region of strong current along the separator itself. The value of $\eta$ is chosen such that it is larger than the numerical diffusivity. For reference, the current sheet dimensions are approximately $1\times5\times134$ (width $\times$ breadth $\times$ height), for a critical current $j_{\rm{crit}}=18$.

\begin{figure*}
 \centering
 \sbox{\bigleftbox}{%
 \begin{minipage}[b]{.47\textwidth}
  \centering
  \vspace*{\fill}
  \subfloat[Electron initial positions ($-20<x,y<20$\unit{km})]
  {\label{subfig:Einiorbsl4}\resizebox{\textwidth}{!}{\includegraphics{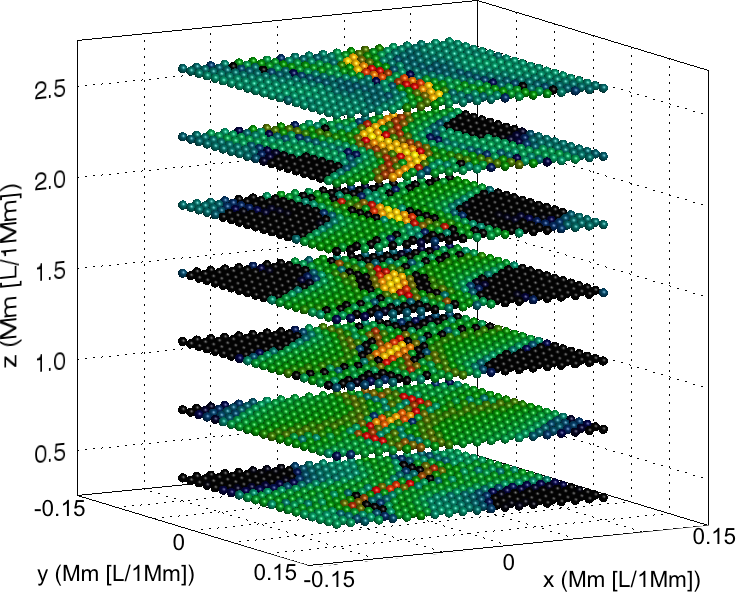}}}\\
  \vspace*{1em}
  \renewcommand{\thesubfigure}{c}
  {\resizebox{\textwidth}{!}{\includegraphics{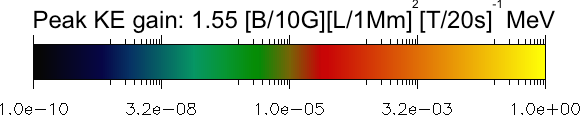}}}\\
  \subfloat[Proton initial positions ($-20<x,y<20$\unit{km})]{\label{subfig:Piniorbsl4}\resizebox{\textwidth}{!}{\includegraphics{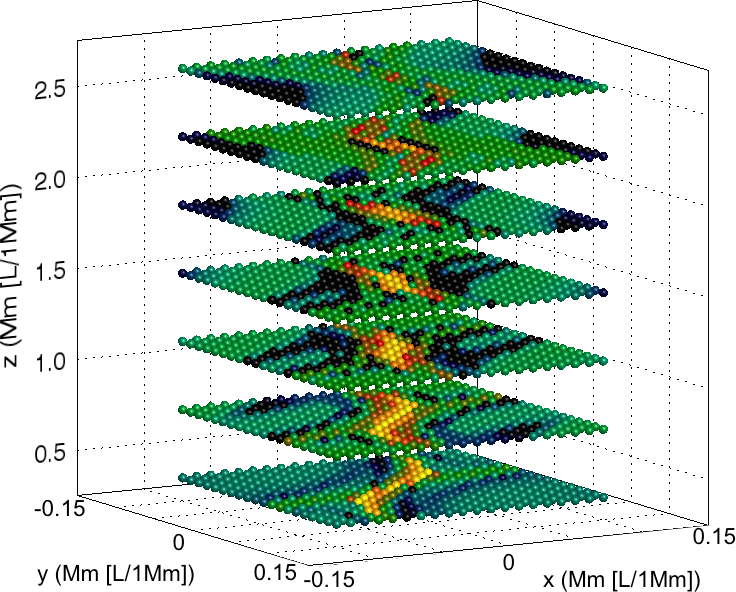}}}
 \end{minipage}%
 }\usebox{\bigleftbox}%
 \begin{minipage}[b][\ht\bigleftbox][s]{.49\textwidth}
  \centering
    \renewcommand{\thesubfigure}{b}
   \subfloat[Electron final positions$\qquad\qquad$] {\label{subfig:Efinalorbsl4}\resizebox{\textwidth}{!}{\includegraphics{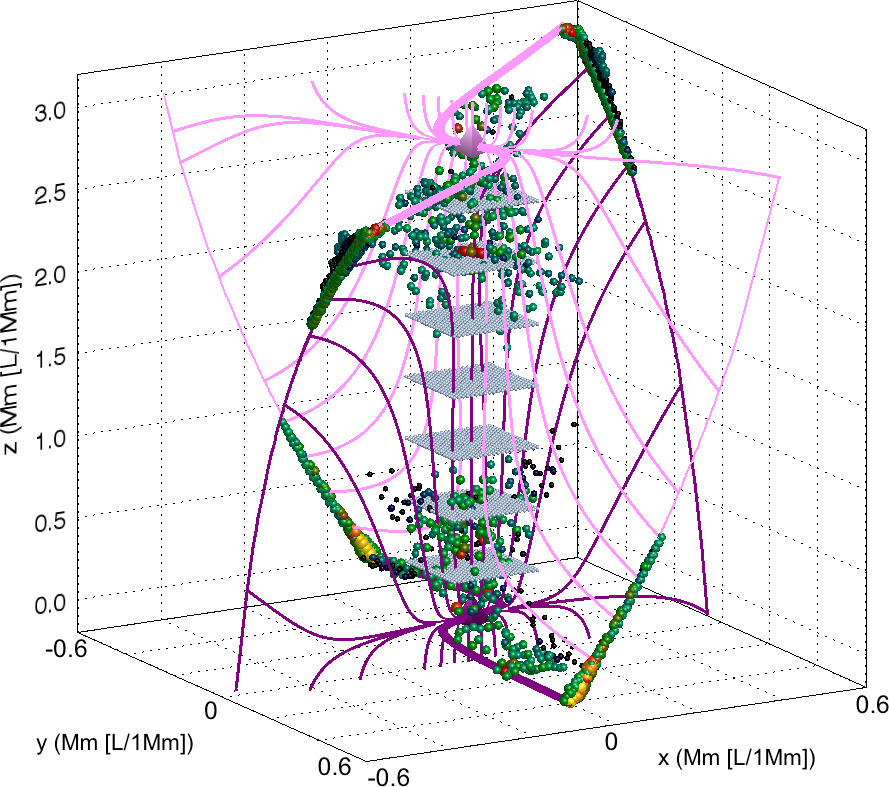}}}\\
     \renewcommand{\thesubfigure}{d}
   \subfloat[Proton final positions $\qquad\qquad$] {\label{subfig:Pfinalorbsl4}\resizebox{\textwidth}{!}{\includegraphics{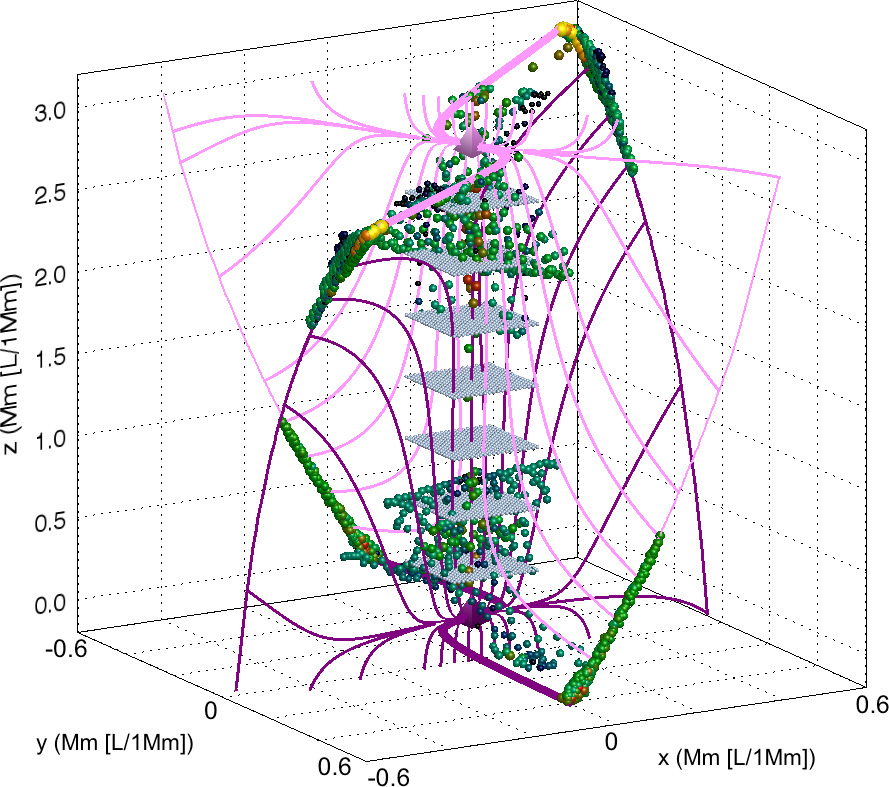}}}
 \end{minipage}
 \caption{Numerical MHD separator model 
 starting from MHS1; Electron \protect\subref{subfig:Einiorbsl4} initial and \protect\subref{subfig:Efinalorbsl4} final positions with initial kinetic energy $20$\unit{eV} and initial pitch angle $45\degr$. \protect\subref{subfig:Piniorbsl4} and \protect\subref{subfig:Pfinalorbsl4} are the same, but for protons. The colour of each position identifies the (scaled) peak kinetic energy gained during each orbit (see colour bar). For \protect\subref{subfig:Efinalorbsl4} and \protect\subref{subfig:Pfinalorbsl4}, the grey orbs indicate the initial position grid, while pink/purple lines indicate interpolated field lines in the fan planes associated with the upper/lower null (indicated by trapezoids) and thick pink/purple lines indicate the corresponding null spines. To aid visualisation, the orb size for each final position has been enhanced in proportion to the final energy. For reference, $\bscl=10$\unit{G}, $\lscl=1$\unit{Mm}, and $\tscl=20$\unit{s} for these orbit simulations.} 
 \label{fig:l4}
\end{figure*}

Like the results described in Sec.~\ref{subsec:scaling}, the particle simulations in this environment begin with a uniform grid of initial conditions. We initialise 2800 particle orbits, each with an initial pitch angle of $45\degr$ and an initial energy of $20$\unit{eV} in a grid centred on the separator. 
The grid consists of 7 planes at different vertical heights ($z=0.375, 0.75,1.125,1.50,1.875,2.25,2.625$\unit{Mm}), while each plane contains $20^2$ positions, evenly spaced from $-0.1,0.1$\unit{Mm} in $x/y$. The orbit calculations are terminated if the particle leaves through certain spatial boundaries; for the present investigation, we define these boundaries at $-0.6$\unit{Mm} and $0.6$\unit{Mm} in $x,y$ and at $0$\unit{Mm} and $3$\unit{Mm} in $z$. As before, we simulate orbits of both electrons and protons; the results of this survey of both species behaviour can be seen in Fig.~\ref{fig:l4}.

Our analysis begins with the distribution of peak energy gain for electrons (Fig.~\ref{subfig:Einiorbsl4}) and protons (Fig.~\ref{subfig:Piniorbsl4}) illustrated on plots of the initial particle positions. The distribution of final energies relative to the initial particle positions closely reflects the electric field configuration in the MHD model snapshot, in agreement with \citet{paper:Threlfalletal2015}. Regions where large energy gains are recovered are strongly associated with regions of current at/above the current threshold ($j_{\rm{crit}}$) in the model. The same is also true for the analytical separator model results, in Figs.~\ref{subfig:JT2015einipos} \&~\ref{subfig:JT2015pinipos}, where the distributions are strongly related to the electric field configuration.

Figures~\ref{subfig:Einiorbsl4} \&~\ref{subfig:Piniorbsl4} also illustrate a vertical asymmetry in energy gains, similar to that seen in the analytical case. Electron acceleration is more strongly associated with the top of the reconnection region compared to protons, which are typically accelerated strongly from the bottom. A related asymmetry is readily apparent in the analytical case (Figs.~\ref{subfig:JT2015einipos} \&~\ref{subfig:JT2015pinipos}), but in the opposite sense, because the electric field configuration in \citet{paper:Threlfalletal2015} is aligned anti-parallel to the separator, rather than parallel to the separator as it is in the MHD model. Electrons and protons are still capable of significant acceleration from part way along the separator, but spend more time in the electric field when originating at a particular end of the reconnection region; these locations are determined by the sign of $E_{||}$ within the diffusion region. 

An initial study of the width of the regions where large energy gains are achieved in Fig.~\ref{subfig:Einiorbsl4} might suggest that the acceleration region is much smaller than that seen in the analytical experiment (e.g. Fig.~\ref{subfig:JT2015einipos}). One must remember, however, that the initial grid seen in Fig.~\ref{subfig:JT2015einipos} is significantly smaller than the relative size of the initial grid seen in Fig.~\ref{subfig:Einiorbsl4}. The grid in Fig.~\ref{subfig:JT2015einipos} was chosen to reflect the width of the reconnection region, which (for the results seen in that image) is approximately 72\unit{m}, for a null separation distance of $10$\unit{Mm}. If the null separation in the MHD experiment was also $10\unit{Mm}$, the current sheet thickness and breadth would approximately be $45\unit{km}$ and $210\unit{km}$, respectively. The MHD experiment includes a much broader reconnection region (compared to its vertical length) than the analytical model. In order to achieve comparable widths, the numerical model would either require a dramatic reduction in non-dimensional reconnection region width or breadth (requiring either a significantly higher resolution or perhaps a new method of defining the reconnection region itself) or a much shorter dimensional separator length (significantly reducing the particle energy ranges recovered).

The peak energy gains seen in Fig.~\ref{fig:l4} are approximately fifteen times larger than those recovered by the analytical model in Fig.~\ref{fig:JT2015}; for the normalising values used in this experiment, the peak kinetic energy gained in the simulation by both electrons and protons was close to $1.5$\unit{MeV}. However, a different choice of dimensional quantities would yield higher peak energies. As an example, an increase in length-scale only by a factor of ten (i.e. a null separation distance of $28.7$\unit{Mm}) would yield a hundred-fold increase in the peak energy gains, allowing particles to potentially achieve $155$\unit{MeV} in energy. Applying the same normalisation as that used in \citet{paper:Threlfalletal2015} ($\bscl =100$\unit{G}, $\lscl =10$\unit{Mm}, and $\tscl =100$\unit{s}) would scale up the energy distribution of these particles further, achieving a new peak kinetic energy around $310$\unit{MeV}. Once again, we may use the results presented as a guide, to define an upper bound of particle energy gains for a separator of a certain length defined by, for example, observations (which suggest that lengths $\geq10\unit{Mm}$ are relatively common within the solar atmosphere, as discussed in Sect.~\ref{sec:disc}). In defining such an upper bound, it should be borne in mind that the reconnection region size and electric field in the numerical MHD model is much broader and stronger than that specified in the analytical model due to numerical constraints.

The final particle orbit positions, shown in Fig.~\ref{subfig:Efinalorbsl4} for electrons and Fig.~\ref{subfig:Pfinalorbsl4} for protons show a degree of correlation with the particle paths found from the analytical model. In Fig.~\ref{subfig:JT2015eorbs}, we see that electrons are accelerated to two specific sites located along one of the fan planes, close to the spine of the upper null. Meanwhile, Fig.~\ref{subfig:JT2015porbs} highlights proton acceleration to two different sites, along the opposite fan plane, close to the spine of the lower null. Figures~\ref{subfig:Efinalorbsl4} \&~\ref{subfig:Pfinalorbsl4} recover similar behaviour; electron and proton orbits are essentially associated with one fan-plane or the other, depending on particle species. In particular, high energy particles are directly accelerated along the fan planes themselves. The main difference between orbits in the analytical and numerical cases is the diffuse nature of the acceleration sites. The final positions of high energy electron orbits (in Fig.~\ref{subfig:Efinalorbsl4}) and proton orbits (in Fig.~\ref{subfig:Pfinalorbsl4}) are more broadly distributed along their respective fan planes in the numerical separator reconnection model. In the analytical model, impact sites are more highly focussed and are directly aligned with the spines of the relevant null. 

Finally, we must confirm that our use of the guiding centre approximation is justified. Our approach always recovers small and finite values for the maximum gyro-radius. This implies that orbit trajectories do not pass close to either of the nulls in the system, where the guiding centre approximation will ultimately break down. More specifically, for the orbits shown in Fig.~\ref{fig:l4}, the largest electron gyro-radius recorded is approximately $0.64$\unit{m}, while the largest proton gyro-radius recorded is $21.1$\unit{m}; both of these values are much smaller than the normalising length-scale of $1$\unit{Mm} and the size of the grid scale used in the MHD simulations ($3.9\unit{km}$ in $x,y$, $6.5\unit{km}$ in z). As the grid scale of the MHD simulations is larger than the largest gyro radii values seen in this study, this implies that the minimum gradient length scales must be large compared to the gyro radii. Meanwhile, the largest electron gyro-period in the experiment was $1.07\times10^{-7}\unit{s}$, while the equivalent proton value approached $1.29\times10^{-4}\unit{s}$. Individual snapshots in the MHD experiments were taken approximately $0.1\unit{s}$ apart (at a normalising time-scale of $20\unit{s}$) and hence both the length and time-scales involved in the MHD experiment are much greater than those used in the guiding centre approximation approach used here.

The orbits in the analytical case, shown in Fig.~\ref{fig:JT2015}, record a maximum gyro-radius of $4.8\times10^{-3}$\unit{m} for electrons and $0.131$\unit{m} for protons and a maximum electron/proton gyro-period of approximately $1.25\times10^{-8}$/$9.5\times10^{-6}\unit{s}$ respectively. 
As an analytical model its time dependence is continuous rather than discrete, but by using length and time-scales of $\lscl=10\unit{Mm}$, $\tscl=100\unit{s}$ the evolution of the system is much slower than the maximum gyro-period and hence our guiding centre approach is still valid. It is also worth remembering that these orbit calculations are, in principle, scalable; scaling lengths up/down (as discussed earlier) also scales the peak gyro-radius values for these orbit calculations. 

\section{Time variation}\label{sec:time}
\begin{figure}[t]
\includegraphics[width=0.49\textwidth]{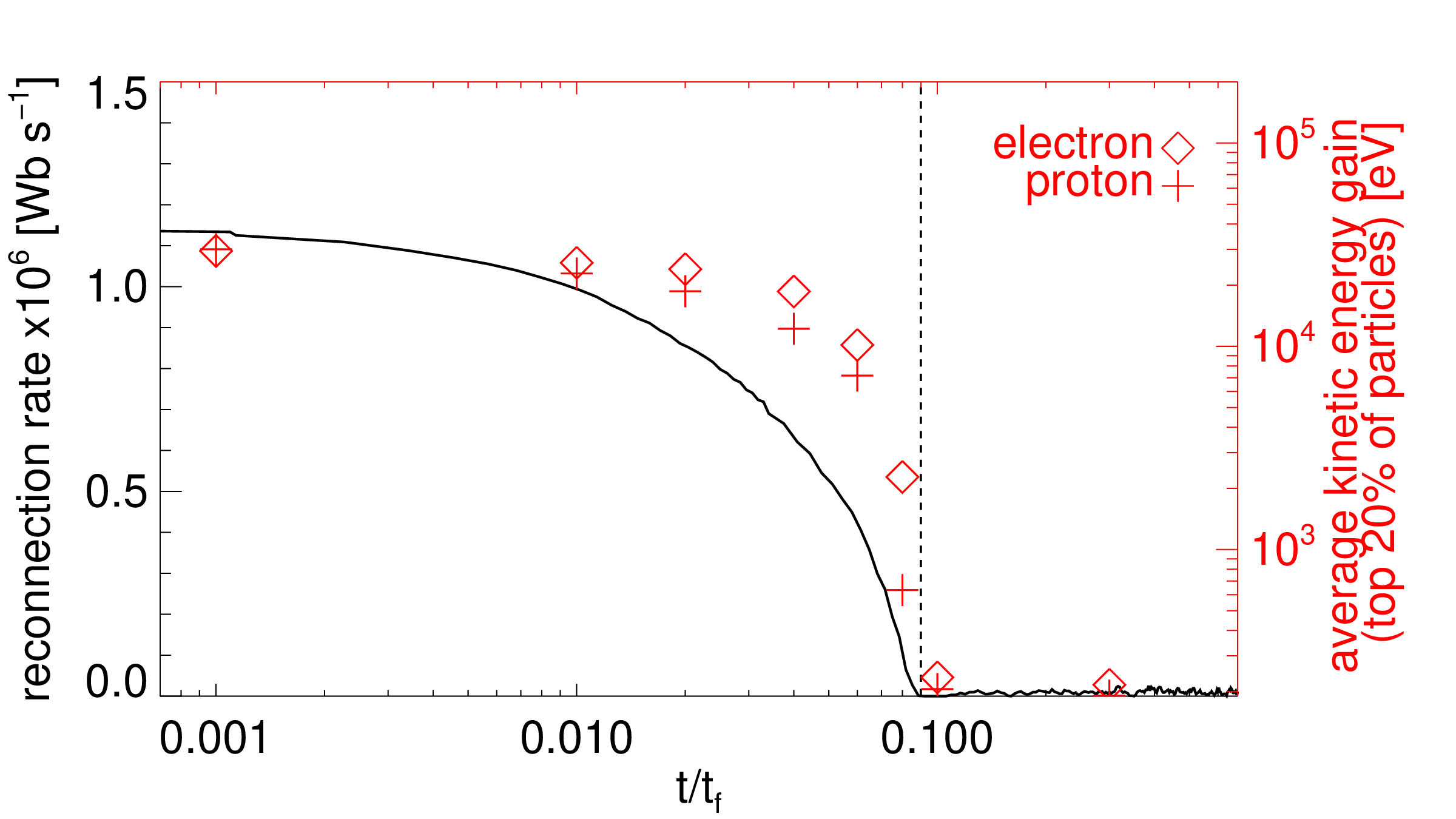} 
\caption{Time evolution of the reconnection rate in the numerical separator reconnection experiment detailed in \citet{paper:Stevensonetal2015}, shown in black, overlaid with average values of the top $20\%$ of the range of kinetic energies gained by particle orbits (electron and proton, see legend) at several snapshots of the reconnection experiment (in red). Several of the particle orbit surveys performed at these snapshots are studied in further detail in Fig.~\ref{fig:tdep}. Times are stated relative to a single fast magneto-acoustic wave crossing time along the original separator, while the reconnection rate is calculated according to the dimensions applied to the MHS2 numerical experiment, ($\bscl =10$\unit{G}, $\lscl =1$\unit{Mm}, and $\tscl =20$\unit{s}).}
\label{fig:tdep_key}
\end{figure}

Up to this point, we have only considered a MHD separator reconnection event when the reconnection rate is close to the peak value. However, as discussed in \citep{inpress:StevensonParnell2015b}, the current sheet which has been built up along the separator dissipates over time. In order to build up a complete picture of particle orbit behaviour during an entire reconnection event, we will now use several different snapshots from the second experiment, MHS2 \citep[detailed in][and previously discussed in Sect.~\ref{subsec:MHDmodel}]{inpress:StevensonParnell2015b} in order to establish how the behaviour discussed in Sect.~\ref{subsec:compy} changes as the reconnection event progresses. As detailed in Table~\ref{table:sepdata}, this second experiment models a high-beta plasma along a shorter separator (non-dimensional length $1.18$). We also retain the normalisation used in the first experiment; once again $\lscl=1$\unit{Mm}, $\bscl=10$\unit{G}, and $\tscl=20$\unit{s} and hence we model a separator of length $1.18$\unit{Mm}.

We use several frames from MHS2 as the basis for our particle orbit analysis, again studying both protons and electrons. One can identify two distinct phases of the experiment, based on the reconnection rate shown in Fig.~\ref{fig:tdep_key}, recorded in terms of the time taken by a fast magneto-acoustic wave to cross the original separator ($t_f$). Reconnection starts instantly when the experiment starts [$t=0t_f$], with a peak reconnection rate of $1.14\times10^6\unit{Wb\,s^{-1}}$($=1.14\times10^6\unit{V}$). This is the start of phase 1, after which the reconnection rate rapidly falls from the peak value to zero [at $t=0.090t_f$, denoted with a dashed line in Fig.~\ref{fig:tdep_key}]. The second phase then begins, in which the steady flows left in the wake of the main reconnection phase cause weak, sporadic reconnection until the end of the experiment [$t=0.5t_f$]. This weak sporadic reconnection typically keeps the reconnection rate above zero. The time-scale considered here might appear to suggest an extremely rapid reconnection, which lasts only a very small fraction of $t_f$. However, $t_f$ represents the time taken for a fast-mode wave to cross the length of the separator. If instead we were to consider the time taken for a fast mode wave to cross the width of the current layer, (e.g. $t_f^{\rm cl}$), we would correspondingly scale up the times presented in Fig.~\ref{fig:tdep_key} (e.g. $t_f=20t_f^{\rm cl}$ in MHS2, based on the ratio of separator length to current layer width).

To gain insight into the particle behaviour, we perform surveys of electron and proton orbits with initial conditions similar to those outlined in Sect.~\ref{subsec:compy}. We simulate 1792 particle orbits with an initial pitch angle of $45\degr$ and an initial energy of $20$\unit{eV}, which begin in 7 $z$-planes (at 0.2,0.3,0.4,0.5,0.6,0.7,0.8\unit{Mm}) each consisting of $16^2$ particles, evenly spaced from $[-0.2,0.2]$\unit{Mm} in $x/y$. The orbit calculations are terminated if the orbit passes beyond $[-0.6,0.6]$\unit{Mm} in $x/y$ and $[0,1]$\unit{Mm} in $z$. 

We highlight the behaviour of both electrons and protons by comparing the average kinetic energy of the highest $20\%$ of peak kinetic energies recorded using $8$ snapshots at various stages of the experiment against the reconnection rate in Fig.~\ref{fig:tdep_key}. In comparing the average of the peak energies, we aim to capture both a sense of global trends and the behaviour of the particles which have undergone the most acceleration. In Fig.~\ref{fig:tdep_key}, we show that the amount of acceleration of the most highly energised electrons and protons is closely related to the reconnection rate; during phase 1, average peak kinetic energies decrease from $\sim30$\unit{keV} to a much lower value ($\sim200$\unit{eV}) when the reconnection rate becomes close to zero. They remain at this low (thermal) value during phase 2. Figure~\ref{fig:tdep_key} also shows that, in general, the electrons and protons behave in the same manner, but that the average peak kinetic energy of protons decreases, approaching the thermal value faster than the electrons as time progresses.

\begin{figure*}[t]
  \centering
{\includegraphics[width=0.6\textwidth]{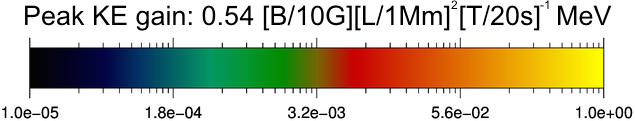}}\\
 \subfloat[$t=0t_f$; electrons]{\label{subfig:td0e}\includegraphics[width=0.31\textwidth]{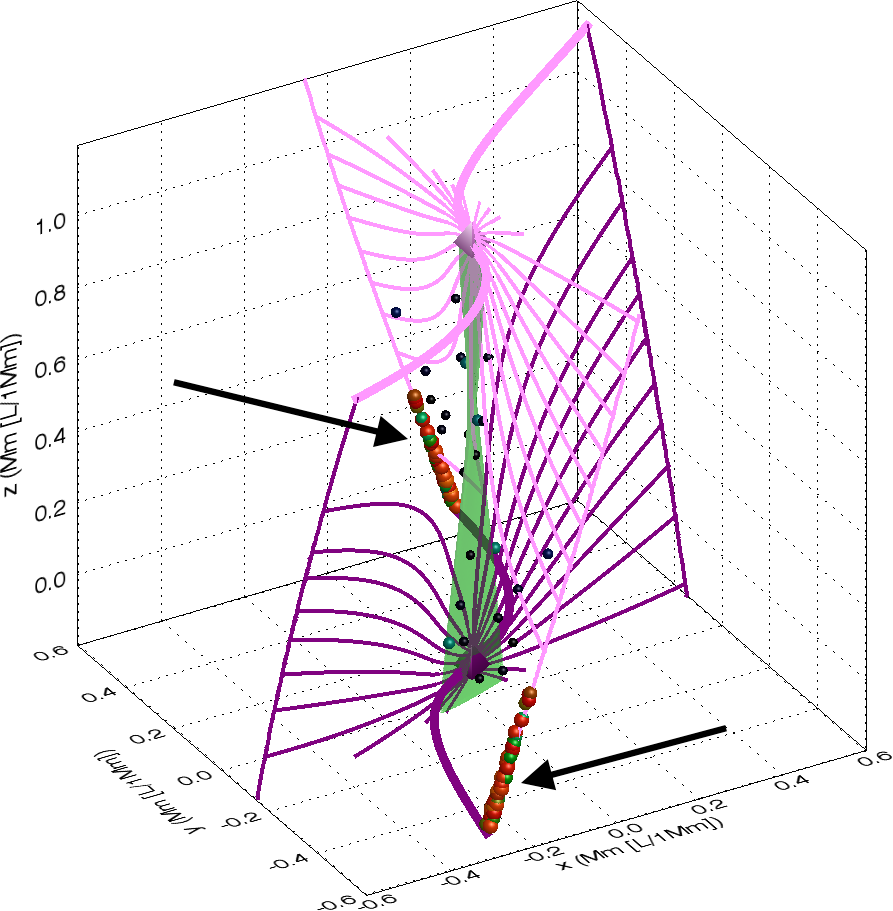}}
 \subfloat[$t=0.06t_f$; electrons]{\label{subfig:td2e}\includegraphics[width=0.31\textwidth]{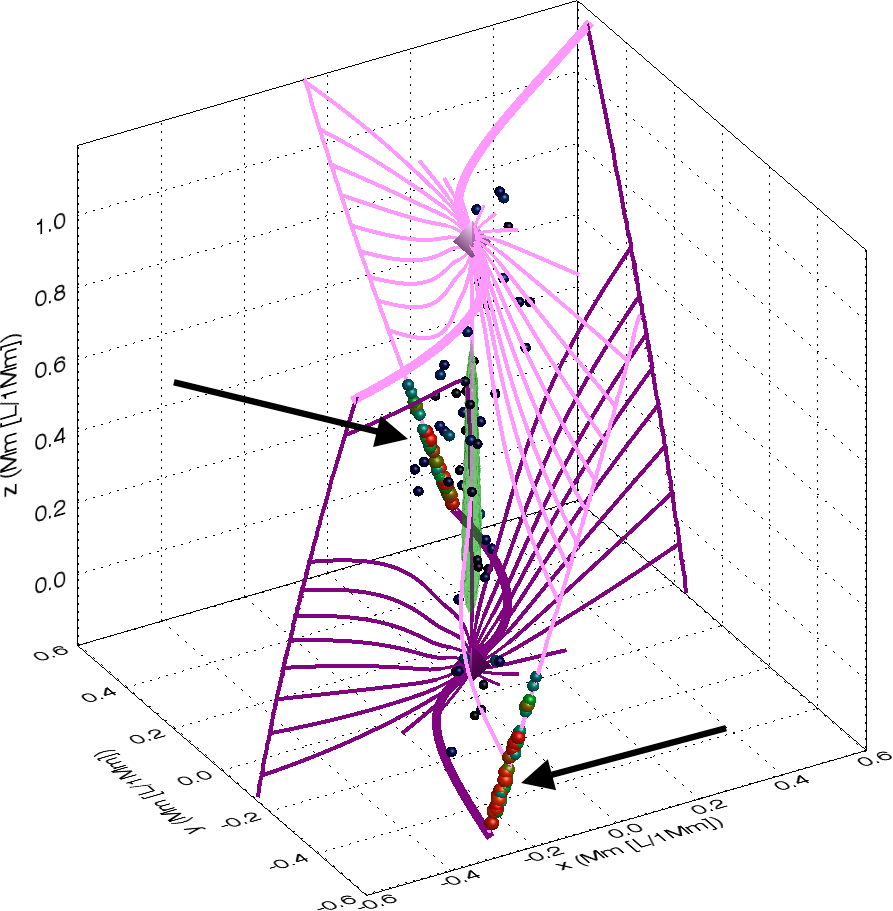}}
 \subfloat[$t=0.1t_f$; electrons]{\label{subfig:td4e}\includegraphics[width=0.31\textwidth]{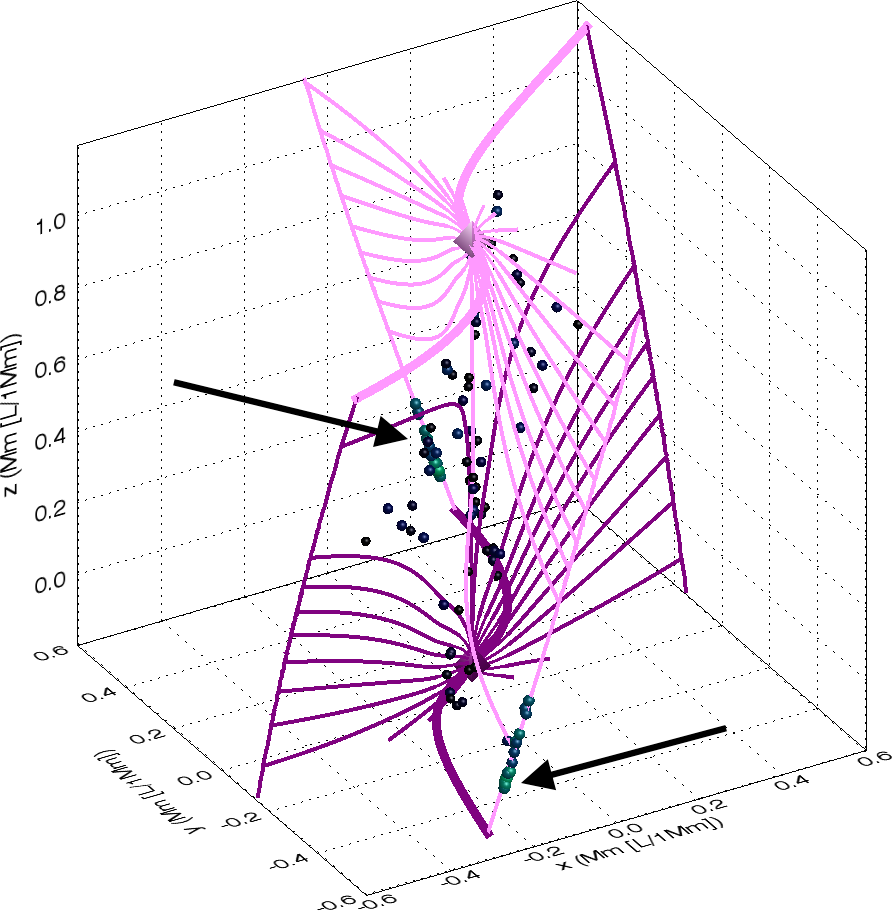}}\\
 \subfloat[$t=0t_f$; protons]{\label{subfig:td0p}\includegraphics[width=0.31\textwidth]{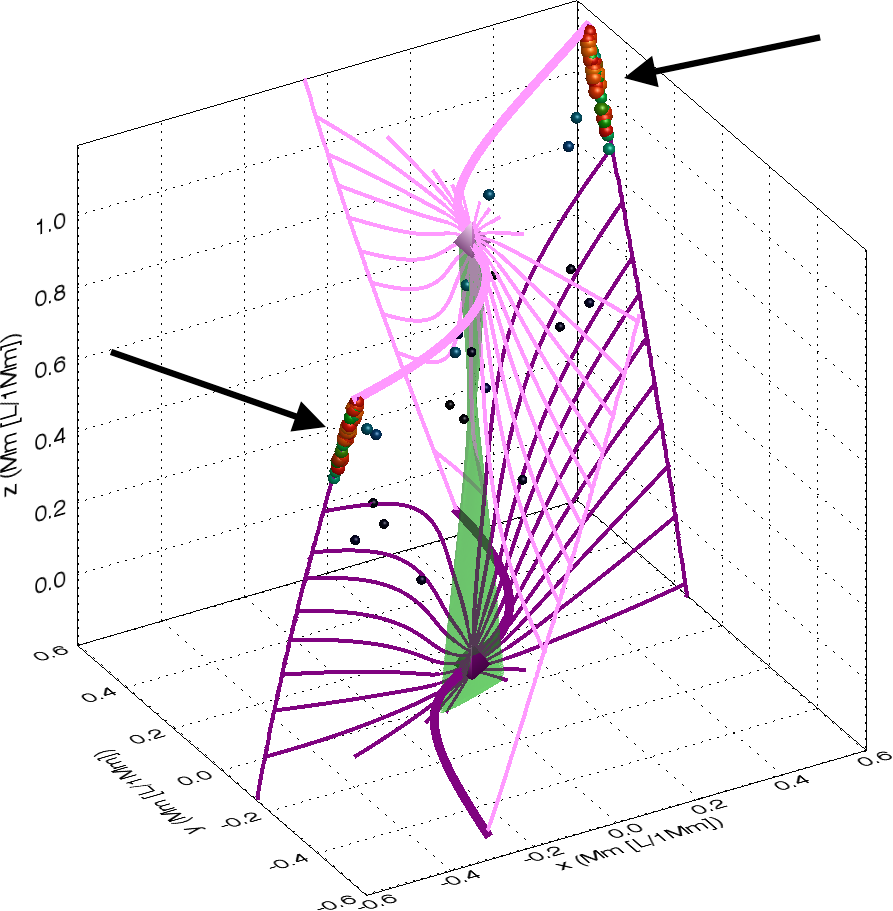}}
 \subfloat[$t=0.06t_f$; protons]{\label{subfig:td2p}\includegraphics[width=0.31\textwidth]{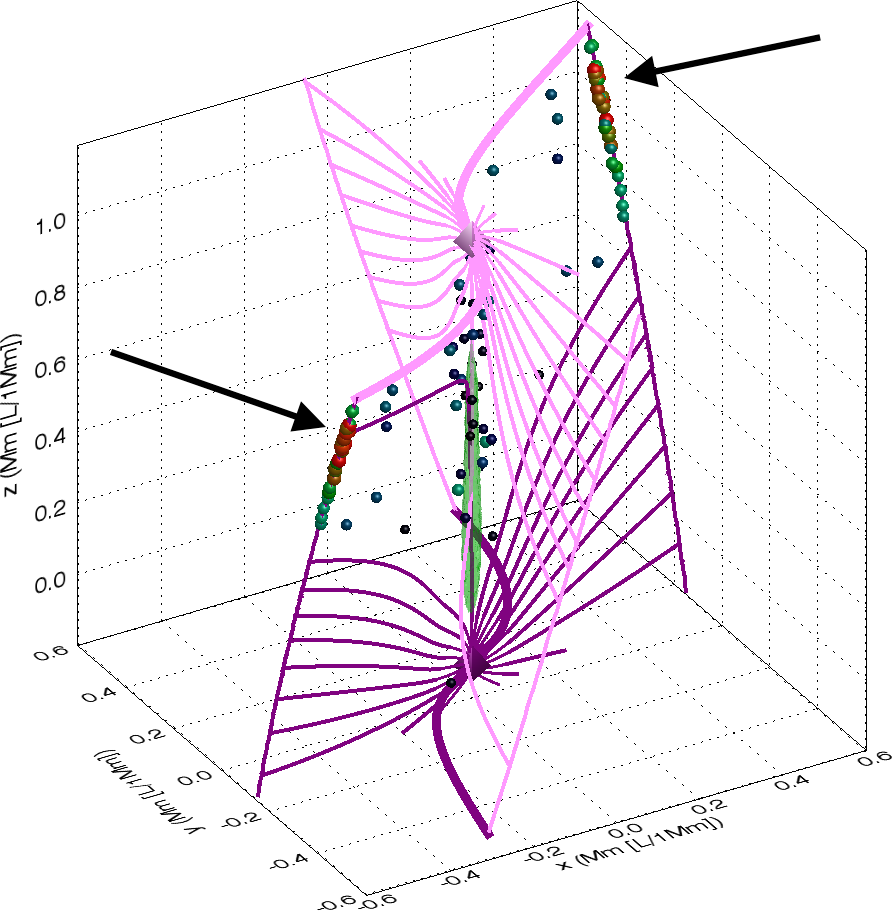}} 
 \subfloat[$t=0.1t_f$; protons]{\label{subfig:td4p}\includegraphics[width=0.31\textwidth]{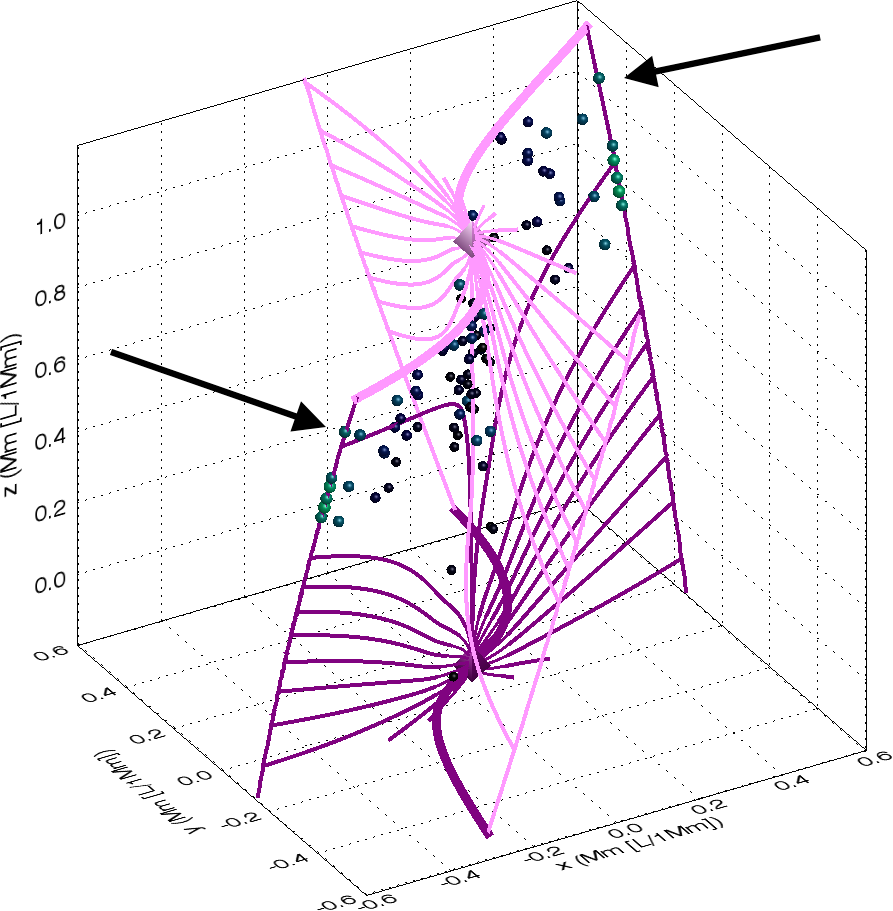}}
 \caption{Numerical MHD separator model (MHS2); 
  Final electron and proton positions coloured by particle energy gain, based on different snapshots during the MHD experiment. Each row depicts a different snapshot, for electron (top) and proton (bottom) positions and each is overlaid with the corresponding isosurface of critical current, $j_{\rm{crit}}$ (in green) for that snapshot. Black arrows indicate typical final positions of accelerated particles, as a guide; these locations most strongly reflect changes in particle behaviour as experiment evolves. Topological features seen here are coloured in an identical manner to those identified in Fig.~\ref{fig:sepcartoon} and previous experiments. For reference, $\bscl=10$\unit{G}, $\lscl=1$\unit{Mm}, and $\tscl=20$\unit{s} for these orbit simulations.}
 \label{fig:tdep}
\end{figure*}

In an effort to better understand this result, we display the electron and proton final positions (colour-coded by energy) at three different stages of the reconnection experiment, MHS2, in Fig.~\ref{fig:tdep}. The particular snapshots were chosen to represent particle behaviour when the reconnection rate was close to its peak value ($t=0t_f$, Figs.~\ref{subfig:td0e} \&~\ref{subfig:td0p}), when the reconnection rate had fallen to approximately half its peak value ($t=0.06t_f$, Figs.~\ref{subfig:td2e} \&~\ref{subfig:td2p}) and after the reconnection rate achieved its minimum value ($t=0.1t_f$, Figs.~\ref{subfig:td4e} \&~\ref{subfig:td4p}). Figure~\ref{fig:tdep} also shows the evolution of both topological features and the current sheet at the value of the critical current threshold above which the anomalous resistivity is triggered. For clarity, we only include orbits which, for the present normalisation, gain more than $10$\unit{eV} in energy, in order to focus purely on the particles which enter the reconnection region.

Several features are readily apparent from Fig.~\ref{fig:tdep}. Firstly, the isosurface of critical current diminishes with time; in effect, this shows the volume of the reconnection region decreases over time, until (at $t=0.09t_f$) it is no longer present, meaning that the current located at the separator has now dropped below the critical level, $j_{crit}=10$. As a result, fewer particles are visible in later stages of the experiment and of these, many no longer achieve high energy gains. The sites of accelerated particles indicated by black arrows in Fig.~\ref{fig:tdep} clearly demonstrate this result. 

When the reconnection rate is close to its peak value, large numbers of electron/proton orbits are found along the separatrix surfaces (in Fig.~\ref{subfig:td0e}/\ref{subfig:td0p}), close to the spine of lower/upper null. As the diffusion region diminishes, fewer electrons/protons are accelerated to the same sites (see e.g. Fig.~\ref{subfig:td2e}/\ref{subfig:td2p}); once the diffusion region has virtually disappeared, almost no electrons/protons are accelerated to high energies and very few reach these locations with any energy (see e.g. Fig.~\ref{subfig:td4e}/\ref{subfig:td4p}). In phase 2 ($t\geq0.09t_f$), the majority of orbits gain only very small amounts of energy (for the normalising values chosen in this experiment, this corresponds to increases of $50-500$\unit{eV}) and remain in the centre of the domain. This is because, as time progresses, both the strength and extent of the reconnection electric field diminishes, reducing both the number of orbits which interact with the reconnection region and the length and strength of the field-aligned potential which causes the acceleration.   

Looking again more closely at the sites of accelerated particles indicated by black arrows in Fig.~\ref{fig:tdep}, we note that the locations reached by those particles which are accelerated move in time; as time progresses, acceleration sites move down the fan planes away from the spines. The field lines which lie close to the separator (and close to the nulls, before extending out into the domain close to the spines of each null) are reconnected first. Field lines which lie further away from the separator are then brought in and are reconnected. These additional field lines are not (initially) close to the separator and neither do they closely approach either null. Instead they are tethered (outside the reconnection region) within the same fan planes just further from the null-spines. Over time, field lines that originally lie progressively further out from the separator are brought in and reconnected, leading to the accelerated orbit impact sites moving away from the spines along the intersection of the fan plane with the domain boundary.

In comparison with earlier experiments, another difference recovered here are the values of peak kinetic energies gained by the individual orbits. The peak kinetic energy gain is smaller than the equivalent value recorded in Sec.~\ref{subsec:compy}; this can clearly be seen in the colour scale of Fig.~\ref{fig:tdep} which is scaled by $0.54$\unit{MeV}, rather than $1.55$\unit{MeV} as shown in Fig.~\ref{fig:l4}. 
This change arises from both a reduction in the peak/mean electric field, but, more importantly, a reduction in non-dimensional separator length, from $2.87$ to $1.18$ units. In Sec.~\ref{subsec:scaling}, we discuss how changes in the dimensional scaling, $\lscl$, affect the peak energy that a particle orbit may achieve. In this instance, a reduction in the non-dimensional (as opposed to the dimensional) separator length of $58\%(\approx1/\sqrt{3})$ has the same effect, resulting in the peak kinetic energies gained by the particles reducing to a third of that seen in Sec.~\ref{subsec:scaling} (if we hold $\lscl=1$\unit{Mm} fixed). Again, this ($L^2$) decrease results from a reduction in size of parallel electric field strengths {\emph{and}} the length over which the potential difference is applied.

At the reference values of normalising field strength ($\bscl=10$\unit{G}), length-scale ($\lscl=1$\unit{Mm}), and time-scale ($\tscl=20$\unit{s}), the peak kinetic energy gained using any snapshot for either particle species was achieved by a proton, which gained $0.542$\unit{MeV} at $t=0t_f$, when the reconnection rate was close to the peak value. The peak energy gained by an electron in the same snapshot was $0.491$\unit{MeV}. Both species often produce similar maximum energy gains, but asymmetries in the current sheet extent and the use of a discrete grid of initial positions mean that no consistent pattern emerges regarding the absolute maximum peak energy and whether this is always achieved by an electron or proton. By taking the average over the most highly energised electrons and protons at any given snapshot (shown in Fig.~\ref{fig:tdep_key}), it becomes clear that, on average, electrons are able to gain more energy than protons, apart from times when the reconnection rate is near the peak value, or near zero, where the species energies are typically well matched. Instances of weak acceleration preferentially accelerate electrons due to the electron/proton mass difference; electrons are lighter and more readily accelerated than protons, unless the force accelerating them is extremely strong or near-negligible.

\section{Discussion}\label{sec:disc}
In this work, we have studied the behaviour of particles in the vicinity of separators which undergo magnetic reconnection, using two different time-dependent models of the separator reconnection event itself. Both the analytical model results (described in Sec.~\ref{subsec:scaling}) and the numerical MHD simulation snapshots (including examples of low/high beta plasmas and longer/shorter separator examples described in Sec.~\ref{subsec:compy}-\ref{sec:time}) result in a significant fraction of electrons or protons being rapidly accelerated to high (non-thermal) energies.

Both models tackle the problem in different ways and yet there is a high degree of similarity between the particle orbit results. The analytical model of \citet{paper:Wilmot-SmithHornig2011} only describes the evolution of electric and magnetic fields in a manner which satisfies Faraday's Law and was designed to study the creation and evolution of topology in multiple separators. In doing so, it is able to describe length-scales in the problem in a way that the numerical MHD model cannot (without dramatic increases in resolution). Indeed, the aspect ratio of the reconnection region width to separator length is approximately $10^{-6}:1$, compared to the numerical models, whose ratio of reconnection region breadth and thickness to separator length is approximately $4\times10^{-3}:3\times10^{-2}:1$ in MHS1 and $5\times10^{-2}:2\times10^{-1}:1$ in MHS2 when reconnection is initiated. However, the numerical models describe additional behaviour, in the sense that the full set of MHD equations (describing all aspects of plasma behaviour, not simply electric and magnetic fields) are solved. Despite these apparent differences, the resulting particle orbits show that any encounter with the reconnection region typically leads to strong direct acceleration. The amount of acceleration typically increases for trajectories which more closely approach the original separator.

An asymmetry also develops between particle species in both cases; electrons and protons are typically more strongly accelerated from opposite ends of the separator and leave the system via the fan-planes of opposite nulls (close to the opposing null spines, if the trajectory of the orbit is close to the separator itself). These findings \citep[and by extension those seen in][]{paper:Threlfalletal2015} appear to be generic features of particle acceleration at reconnecting magnetic separators. The acceleration of protons and electrons along different sets of field-lines may be regarded as the three-dimensional equivalent of the results of \citet{paper:ZharkovaGordovskyy2004} who considered a two-dimensional current-sheet plus guide-field configuration (however, it has to be noted that the equivalence is not perfect, as particles in the 3D separator models considered here may encounter mirror points outside the reconnection region and return, unlike the earlier 2D current-sheet configuration). The acceleration of electrons and protons to different sites which depend on particle species may also contribute to the lack of alignment of X-ray and $\gamma$-ray features observed during a single solar flare event \citep[e.g.][]{paper:Hurfordetal2003}. 

The particle orbits which result from the numerical models (Sec.~\ref{subsec:compy}-\ref{sec:time}) differ from those seen in the analytical model (Sec.~\ref{subsec:scaling}) in one crucial aspect; in the latter case beams of accelerated particles are strongly aligned with the null spines, while in the former case, such orbits may also be accelerated further away from the spine along the corresponding fan plane. This is likely to be due to the difference in reconnection region sizes at/near the separator. In the numerical case, a broader reconnection region (approximately shaped like a ribbon) is threaded by field-lines (and hence accelerated particle trajectories), which do not all closely approach the separator or either null, but which still remain closely aligned with the fan plane structure (within the ribbon-like reconnection region). This is because the reconnection region is located about the two fan planes which lie very close together and intersect close to the centre of the $x,y$ domain \citep{inpress:StevensonParnell2015b}. In the analytical model, the imposed small cylindrical reconnection region is only threaded by field lines which closely follow the separator and, hence, closely approach both nulls and remain close to the spines of each null. 
Additionally (during reconnection), accelerated particle trajectories (which follow reconnected magnetic field lines) arrive at the simulation boundaries progressively further from the spine along the relevant fan plane. This movement of potential impact sites is caused by additional field lines being brought into the reconnection region over time; such field lines are connected to different locations along the same fan plane. 

These findings are in broad agreement with many features of particle acceleration sites seen by observations \citep[see e.g.][]{review:Fletcheretal2011}. Flare ribbons, typically observed in $H\alpha$ (and also in UV/EUV), are seen to evolve over the course of an event and are known to reflect the magnetic topology of a particular event \citep[specifically in the projection of separatrix locations, see e.g.][]{paper:Mandrinietal1991,paper:Demoulinetal1997}. Our findings also suggest that the extent/size of accelerated particle impact sites seen by HXR/$\gamma$-ray data during solar flares is likely to be a proxy for the size of the reconnection region (if particles which leave the acceleration region are not subjected to further processes which might change their trajectory or energy). At present, the relative size of these sites, compared to the large-scale structures they are often associated with \citep[see e.g.][]{review:Fletcheretal2011}, suggests that the reconnection region at the heart of a solar-flare event is typically relatively small.

By analysing several snapshots of an ongoing separator reconnection experiment, we have studied the particle response to the time-dependent process of magnetic reconnection. The results shown in Sect.~\ref{sec:time} show that the particle orbit energies and the locations to which orbits may be accelerated, are intrinsically tied to the reconnection rate. At instances when the reconnection rate is rapid, such as during the impulsive phase of a flare (corresponding to the dissipation of the built-up current over a large volume), particle orbits are accelerated by a large field aligned potential, leading to a significant particle orbit response.
As the volume reduces over time, this causes a reduction in the field aligned potential experienced by the particle. If the diffusion region is almost entirely dissipated (as it is during phase 2, as described in Sect.~\ref{sec:time}) then little to no acceleration is possible. 

In phase 2 of the MHD experiments, reconnection occurs in small, short, local bursts along the separator. Although these bursts occur sufficiently frequently to produce an almost steady (albeit very slow) average reconnection rate, the reconnection site during these bursts is too small and too short lived to have much of an effect on the particle orbits.
This result implies that knowledge of reconnection rates of observed magnetic reconnection events in both the solar atmosphere and the magnetosphere is fundamental to the determination of the potential of such events to accelerate particles. It also may be an explanation as to why hard X-ray emission from flares is typically seen near the start of a flare and is shorter lived compared to the duration of the soft-X-ray emission lifetime \citep{book:Benz2002}. It is important to note, however, that the above discussion is based on the properties of individual particle orbits and not on the calculation of particle fluxes. Hence, while we can discuss purely morphological features, we are not currently in a position to explain emission spectra and other features.

In our model, particle acceleration is caused by the strength and extent of the local parallel electric field resulting from the (ongoing) reconnection. In displaying our results, we have demonstrated a method through which orbit energy gains may be scaled. This allows for an initial direct comparison between all the separator reconnection models used in this work and, in addition, the application of these results to separators which have been observed and/or inferred through observational data and numerical simulations of the solar atmosphere or the magnetosphere.

For example, observations using SoHO/MDI magnetograms reported by \citet[][]{paper:Closeetal2004} suggest that an average of $\sim7$ separators might be associated with a single null and that the mean separator length in a small patch of quiet Sun might approach $100\unit{Mm}$. This length reduces to $60\unit{Mm}$ and $30\unit{Mm}$ with the inclusion of additional large- and small-scale internetwork features in the magnetogram data. While such separators would be located within a small patch in the low corona, a recent global study of topological features in the extended solar corona has been performed by \citet{paper:Plattenetal2014} using synoptic magnetogram data over several solar cycles. This work found that separators shorter than $500\unit{Mm}$ are the most common length, but (often) found (many) additional examples of separators of up to $1600\unit{Mm}$ in length per Carrington rotation \citep[see Fig.~23 of][for further details]{paper:Plattenetal2014}. Finally, numerical simulations which also study the topology of 3D MHD flux emergence models \citep{paper:Parnelletal2010b} identify many separators, ranging from $\sim0.2-36\unit{Mm}$ in length; the majority of the identified separators are between $17-29\unit{Mm}$ long and (due to an associated integrated parallel electric field along their length) undergo magnetic reconnection. While it is difficult to ascertain the proportion of observed separators which are undergoing reconnection (and indeed over how much of the separator length the reconnection would take place), these findings support the presence of separator reconnection, taking place on various scales throughout the solar corona. 

Scaling our results suggests that, for example, the separators identified in numerical models of \citet{paper:Parnelletal2010b} (which are 2-3/20-30 times longer than that modelled analytically/numerically here) could yield particle energy gains of upto $0.4-0.9\unit{MeV}$/$0.6-1.35\unit{GeV}$ respectively. The separator lengths described by \citet{paper:Parnelletal2010b} are also relatively short compared to examples extrapolated from observations, potentially increasing these upper bounds still further. However, our calculations also assume a field strength of $10\unit{G}$ and a reconnection region which extends over one fifth/the entire separator length. We anticipate reconnection region lengths to be much shorter in reality, particularly for long separators which reach into the high corona, since reconnection requires not just a suitable magnetic field configuration in which to occur, but also appropriate plasma conditions and flows. Such conditions are highly unlikely to be achieved all the way along a global separator extending over hundreds of megameters, which would act to reduce the energy particle orbits might gain from them. However, even this simple calculation shows that reconnecting magnetic separators, both observed in nature and recovered in numerical simulations could readily accelerate particles to high energies.

Finally, it is (at present) unclear how a real plasma might support large scale, monolithic current structures (such as those considered here) for long periods of time. In reality, the acceleration of significant quantities of electrons and protons would necessarily lead to feedback upon the original MHD fields, which would in turn affect the acceleration process in a manner which a combined test-particle/MHD approach cannot currently hope to address. Kinetic (for example Particle-In-Cell, or PIC) methods model the interaction of particles and supporting field structures in a self-consistent manner and therefore naturally include the missing feedback. 
However, a realistic 3D kinetic simulation of the dynamics of large-scale coronal structures such as reconnecting magnetic separators are at present not feasible, as kinetic simulations performed at even smaller scales usually have to be carried out with rescaled parameter values. Even if it would be possible to perform 3D kinetic simulations with such rescaled parameter values, it would make the interpretation of the results for investigating particle acceleration very difficult.

\section{Conclusions and future work}\label{sec:conc}
We have studied the relativistic guiding centre behaviour of test particles in numerical snapshots describing the MHD evolution of a reconnecting magnetic separator \citep[following the work of][]{inpress:StevensonParnell2015b}. We have compared the results to test particle behaviour in an analytical model \citep[studied in][]{paper:Threlfalletal2015} and found that particle orbits may be accelerated to high energies for electrons and protons in both cases. Both models also agree that each particle species is most strongly accelerated from opposite ends of the same separator and result in trajectories which terminate along opposite fan planes (where the choice of end and fan plane is species-dependent). 

Most of the differences in our results can be attributed to the different sizes of the reconnection regions in our MHD models.
The main differences between the models are found in the sizes of the reconnection region; the analytical model is capable of describing far smaller reconnection regions, which leads to orbits whose final trajectories are closely aligned with null-spines. Despite being unable to produce similar reconnection region sizes (at present), the numerical MHD model findings suggest that larger reconnection regions would lead to broader impact sites for highly energised particles in the solar atmosphere (i.e. sites of HXR/$\gamma$-ray production during a flare) if particles are not subjected to any further changes in trajectory or energy upon leaving the region studied here. Under such conditions, one might think of the size of impact sites as a proxy for the size/extent of the reconnection region. However, there is general agreement between the orbits found using both approaches; this implies separator reconnection regions may indeed be locations where particles may be (strongly) accelerated.

In agreement with observations, our time-dependent study clearly shows the impact sites of the accelerated particles moves along the fan plane away from the null spines, as field lines that lie further from the separator are brought in and reconnected. Furthermore, the acceleration impact site changes in size and particle characteristics in response to changes in the reconnection rate.

Several extensions to this work are readily apparent. With separator reconnection environments now established as potential particle accelerators, this type of investigation merits repeating for a `more realistic' environment incorporating reconnecting magnetic separators. Specifically, a separator embedded in a coronal magnetic environment, which undergoes reconnection in a manner more closely associated with a solar flare would allow for particle impact site structure to be examined in greater detail and directly compared with HXR/$\gamma$-ray observations. Multiple separators \citep[i.e. separators that link the same pair of nulls, as discussed, for example, in][]{paper:Parnelletal2010b} have begun to be detected/identified in MHD separator reconnection experiments, particularly at times when the reconnection rate is close to its peak value. Whether such separators (which may either thread the same, or disparate diffusion regions) impact the particle acceleration behaviour identified here remains unknown, but is worthy of future consideration. Further development of the test-particle model, including the incorporation of collisional effects and non-uniform, (physically relevant) initial conditions would provide a significant further contribution to our understanding in this area.

\begin{acknowledgements}
The authors gratefully acknowledge the support of the U.K. Science and Technology Facilities Council [Consolidated Grant ST/K000950/1 and a Doctoral Training Grant ST/I505999/1 (JEHS)]. The research leading to these results has received funding from the European Commission's Seventh Framework Programme FP7 under the grant agreement SHOCK (project number 284515). The MHD computations were carried out on the UKMHD consortium cluster housed at St Andrews and funded by STFC and SRIF.
\end{acknowledgements}

\bibliographystyle{aa}        
\bibliography{JT2015c}          
\end{document}

%% file: math_def.tex
\newcommand{\beq}{ \begin{eqnarray} }
\newcommand{\eeq}{ \end{eqnarray} }

\newcommand{\vpar}{ v_{\parallel} }
\newcommand{\vperp}{ v_{\perp} }
\newcommand{\Eperp}{ E_{\perp} }

\newcommand{\boldnabla}{\mbox{\boldmath$\nabla$}}

\def\grad{\boldnabla}

\def\vscl{{v_{scl}}}
\def\vsclsq{{{v_{scl}}^2}}
\def\lscl{{l_{scl}}}
\def\tscl{{t_{scl}}}
\def\bscl{{b_{scl}}}
\def\escl{{{e_{scl}}}}

\def\KEscl{{{{KE}_{scl}}}}
\def\omscl{{{\Omega_{scl}}}}

\newcommand{\Epar}{E_{\parallel}}
\newcommand{\upar}{u_{\parallel}}
\newcommand{\ue}{{{\bf{u}}_E}}
